\pdfoutput=1

\documentclass[11pt]{article}
\usepackage[dvipsnames]{xcolor}

\usepackage[preprint]{acl}

\usepackage{times}
\usepackage{latexsym}

\usepackage[T1]{fontenc}

\usepackage[utf8]{inputenc}

\usepackage{microtype}

\usepackage{inconsolata}

\usepackage{graphicx}
\usepackage{tabularx}
\usepackage{multirow}
\usepackage{amsfonts}
\usepackage{amsmath}
\usepackage{xspace}
\usepackage{lipsum}
\usepackage{tcolorbox}
\usepackage{adjustbox}
\usepackage{booktabs}
\usepackage{makecell}
\usepackage{url}
\usepackage{enumitem}
\usepackage{pgfplots}
\pgfplotsset{compat=1.18}
\usepackage{listings}
\usepackage{subcaption}
\usepackage{algorithm}
\usepackage{algorithmic}

\title{Documentation Retrieval Improves Planning Language Generation}

\author{Renxiang Wang  \quad Li Zhang \\
  Independent \quad Drexel University \\
  {\tt renxiang428@gmail.com} \quad {\tt harry.zhang@drexel.edu}
}

\begin{document}
\maketitle
\begin{abstract}
Certain strong LLMs have shown promise for zero-shot formal planning by generating planning languages like PDDL. Yet, performance of most open-source models under 50B parameters has been reported to be close to zero due to the low-resource nature of these languages. We significantly improve their performance via a series of lightweight pipelines that integrates documentation retrieval with modular code generation and error refinement. With models like Llama-4-Maverick, our best pipeline improves plan correctness from 0\% to over 80\% on the common BlocksWorld domain. However, while syntactic errors are substantially reduced, semantic errors persist in more challenging domains, revealing fundamental limitations in current models' reasoning capabilities.\footnote{Our code and data can be found at~\url{https://github.com/Nangxxxxx/PDDL-RAG}.}
\end{abstract}


\section{Introduction}

\begin{figure}[t!]
	\includegraphics[scale=0.3]{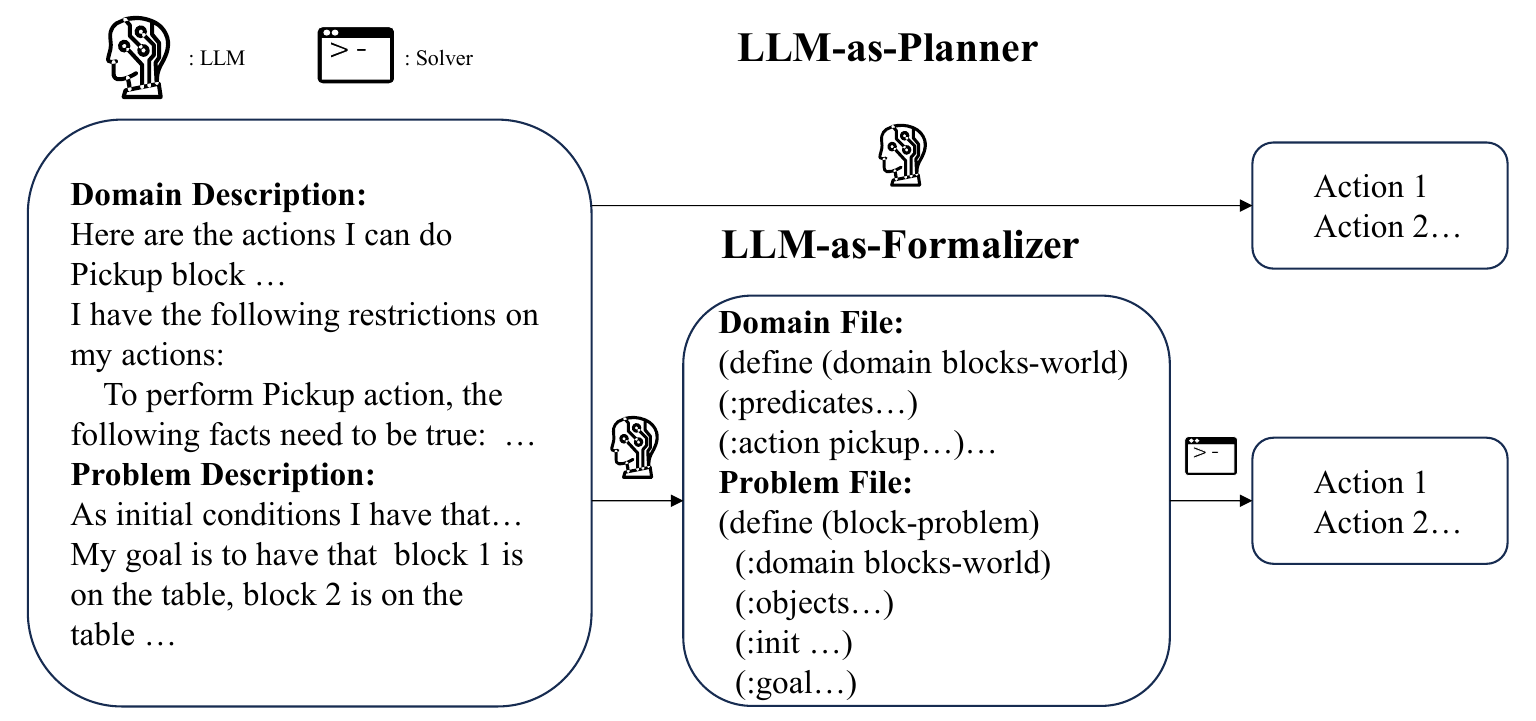}
	\caption{A simplified illustration of LLM-as-Planner and LLM-as-Formalizer on the BlocksWorld domain.}
	\label{fig:Planner_Formalizer}
\end{figure}

Using large language models (LLMs) for planning has garnered significant attention, with two main paradigms as shown in Figure~\ref{fig:Planner_Formalizer}. First, the LLM-as-Planner approach \cite{kambhampati2024llmscantplanhelp,valmeekam2023planbenchextensiblebenchmarkevaluating,stechly2025chainthoughtlessnessanalysiscot,majumder2023clincontinuallylearninglanguage} relies on the reasoning ability of LLMs to directly generate action plans based on descriptions of the environment. In contrast, the LLM-as-Formalizer \cite{tang2024worldcodermodelbasedllmagent,guo2024castlconstraintsspecificationsllm,zhang-etal-2024-pddlego} approach leverages the code generation capability of LLMs to represent the environment in some planning language, which is then passed to a formal solver to derive a plan. Leading to better interpretability and verifiability of the plans, the latter approach has recently gained considerable attention, with Planning Domain Definition Language (PDDL) as one of the predominant formal languages for LLM planning (see the Appendix~\ref{sec:example} for an example of PDDL). 

While LLMs have been shown to somewhat able to generate PDDL, their performance has proven unsatisfactory in realistic and rigorous evaluations \cite{zuo2025planetariumrigorousbenchmarktranslating}. Even state-of-the-art coding LLMs have shown close-to-zero performance as PDDL formalizers on planning benchmarks especially when the model size is less than 100 billion parameters~\cite{huang2025limitlanguagemodelsplanning}, while an array of code generation techniques struggle to improve performance \cite{kagitha2025addressingchallengesplanninglanguage}.
Moreover, training data for low-resource and domain-specific languages like PDDL is extremely limited, making generation even more challenging~\cite{Tarassow2023ThePOA,Joel2024ASOA}. Existing attempts of improvement such as fine-tuning ~\cite{Cassano2023KnowledgeTFA,McKenna2025SyntheticFDA,Giagnorio2025EnhancingCGA} and translation from high-resource languages~\cite{Liu2024SynthesizingPRA} require supervised PDDL data that barely exists. In contrast, retrieval of library documentation~\cite{zhou2023docpromptinggeneratingcoderetrieving,dutta-etal-2024-rar} has proven effective for high-resource languages.


We find that simply providing the documentation to LLMs does not help low-resource PDDL generation. However, we present some novel methods that generate PDDL either modularly or with error refinement, while only retrieving the most relevant documentation. These methods enable a big improvement of PDDL generation performance for models like Llama-4-Scout and Llama-4-Maverick on domains like BlocksWorld, improving correctness from 0\% to 50\%. Moreover, we verify the intuition that documentation significantly reduces syntax errors, but has limited effect on semantic errors. We also present interesting findings that LLMs are more reliant on documentation initially than during error refinement, different models vary in their ability to leverage documentation effectively and that examples are more effective than descriptions in the documentation.

\section{Methodology}
\begin{figure}[t!]
	\includegraphics[scale=0.35]{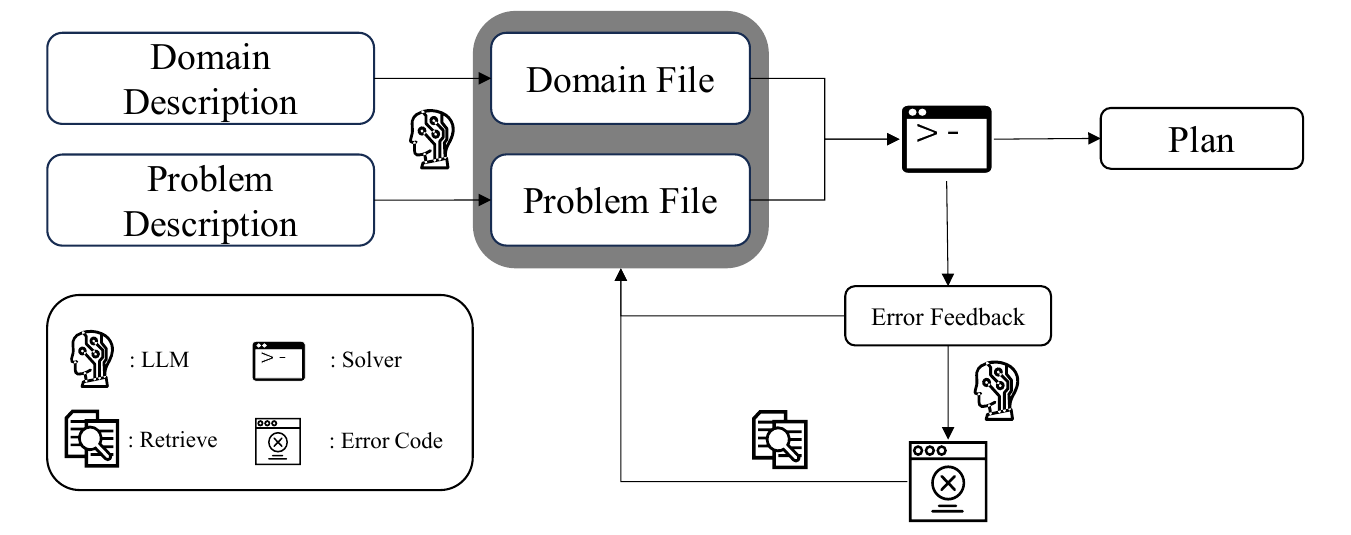}
	\caption{Overview of one of our pipeline that retrieve documents based on error codes located by LLM, and finally using them as hints to correct the code.}
	\label{fig:overview}
\end{figure}

We conduct experiments in text-based simulated planning environments. Each planning problem in the dataset is accompanied by a \textit{domain description} (\texttt{DD}) outlining the environment, and a \textit{problem description} (\texttt{PD}) specifying the task objective.

We begin with the most basic setting, referred to as \textbf{Base}, where a LLM zero-shot generates PDDL code. Given the \texttt{DD} and \texttt{PD} as input, the LLM produces a Domain File (\texttt{DF}) and a Problem File (\texttt{PF}):

\vspace{-8pt}
\begin{equation*}
\mathtt{DF},\ \mathtt{PF} = \textsf{LLM}(\mathtt{DD},\ \mathtt{PD})
\end{equation*}
\vspace{-16pt}

Building upon this, we leverage the PDDL documentation (\texttt{Doc}) during generation. We consider two approaches, \textbf{Once w/ Whole Doc} where the model is given an entire \texttt{Doc} before generating the entire PDDL, and  \textbf{Modular w/ Specific Doc} where the model incrementally generates PDDL code guided by relevant parts of the \texttt{Doc}. Here, we break down the \texttt{DF} structure into types, predicates, actions, etc. and \texttt{PF} structure into initial and goal states. We partition the \texttt{Doc} accordingly. 
\vspace{-8pt}
\begin{equation*}
\mathtt{DF},\ \mathtt{PF} = \textsf{LLM}(\mathtt{DD},\ \mathtt{PD},\ \mathtt{Doc})
\end{equation*}
\vspace{-16pt}

Next, we optionally perform up to three rounds of iterative error correction. We first use a PDDL solver to obtain error feedback:

\vspace{-8pt}
\begin{equation*}
\mathtt{Err\_Feedback} = \textsf{Solver}(\mathtt{DF},\ \mathtt{PF})
\end{equation*}
\vspace{-16pt}

Without the \texttt{Doc}, the standard \textbf{Refinement w/o Doc} directly input the error feedback back to the LLM to re-generate the PDDL:
\vspace{-4pt}
\begin{equation*}
\mathtt{DF},\ \mathtt{PF} = \textsf{LLM}(\mathtt{DF},\ \mathtt{PF},\ \mathtt{Err\_Feedback})
\end{equation*}
\vspace{-16pt}

With the \texttt{Doc}, we attempt to retrieve a specific, helpful part that pertains to the particular error. Using the feedback directly as the query is referred to as \textbf{Refinement w/ Feedback-Retrieved Doc}. Otherwise, we may prompt an LLM to localize the code that caused the error based on the feedback, referred to as \textbf{Refinement w/ Code-Retrieved Doc}.

\vspace{-16pt}
\begin{equation*}
\mathtt{Err\_Code} = \textsf{LLM}(\mathtt{Err\_Feedback})
\end{equation*}
\vspace{-16pt}

In either case, we then retrieve the most relevant documentation snippet using the BM25~\cite{robertson2009probabilistic} retrieval algorithm:

\vspace{-16pt}
\begin{equation*}
\mathtt{Rel\_Doc} = \textsf{BM25}(\mathtt{Err\_Feedback}|\mathtt{Err\_Code})
\end{equation*}
\vspace{-16pt}

Finally, the LLM corrects the code using the retrieved \texttt{Doc}, the \texttt{Error\_Feedback}, and the localized \texttt{Error\_Code} if any.
\vspace{-2pt}
\begin{equation*}
    \begin{split}
        \mathtt{DF},\ \mathtt{PF} = \textsf{LLM}(&\mathtt{DF},\ \mathtt{PF},\ \mathtt{Err\_Feedback}, \\
        &[\mathtt{Err\_Code}],\ \mathtt{Rel\_Doc})
    \end{split}
\end{equation*}
\vspace{-16pt}

The full prompts and the pseudocode are provided in Appendix~\ref{sec:prompt},  and~\ref{sec:pseudocode}. Also we list two examples of how Refinement w/ Code-Retrieved Doc works when facing the wrong PDDL in Appendix~\ref{sec:Error case}

While we only consider PDDL as the planning language in this work following cited works, we also have explored the feasibility of using Satisfiability Modulo Theories (SMT) solvers—specifically Z3, a general-purpose solver for constraint satisfaction planning problems. Following~\citet{hao2025planningrigorgeneralpurposezeroshot}, our evaluation shows that Z3 exhibits suboptimal performance when handling complex planning tasks and is thus not discussed further (see details in Appendix~\ref{sec:Z3}).

\begin{figure*}[t!]
	\includegraphics[scale=0.53]{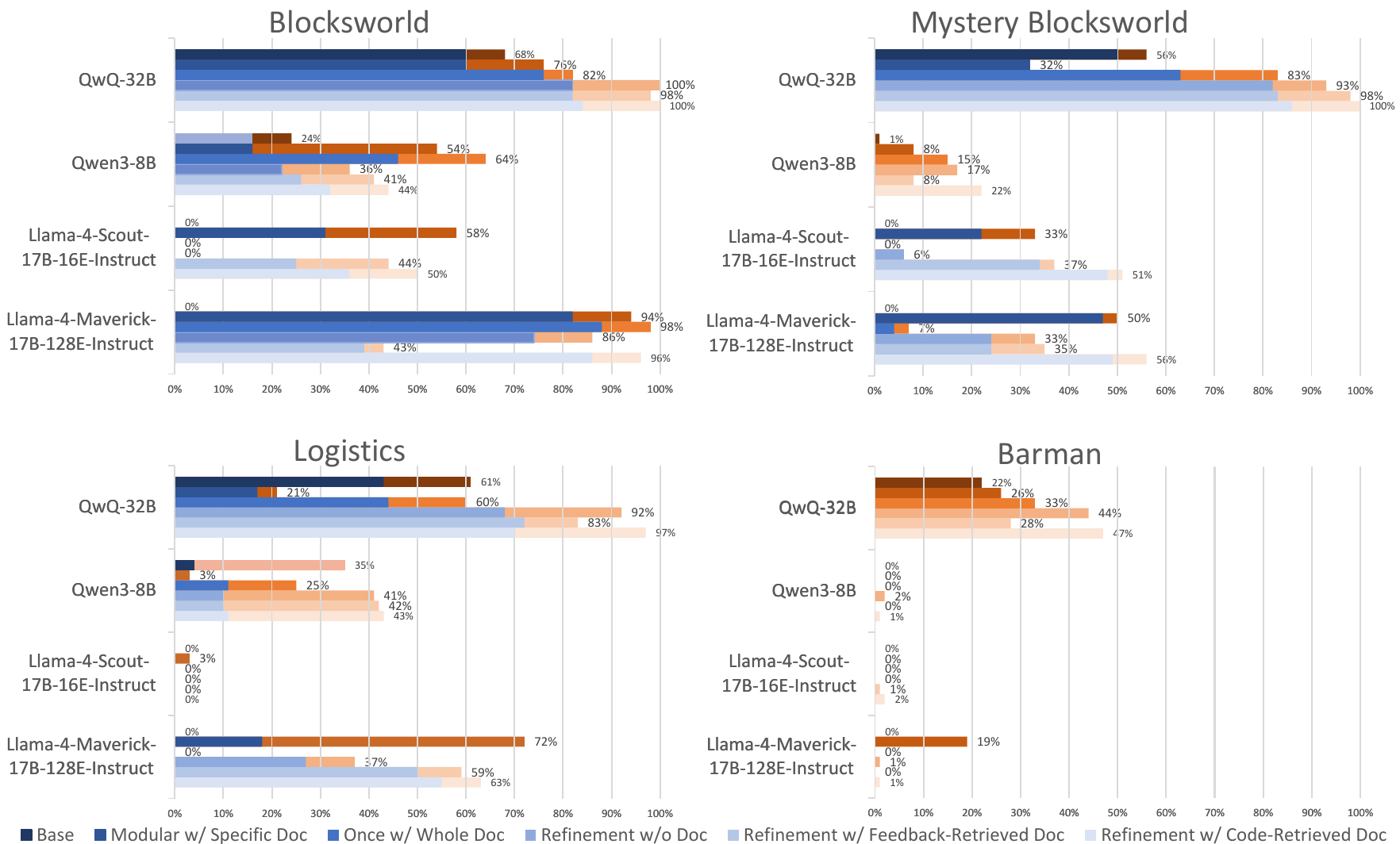}
	\caption{Syntactic accuracy (orange) and semantic accuracy (blue) on various planning domains.}
	\label{fig:Results}
\end{figure*}

\section{Evaluation}

\paragraph{Dataset} To conduct experiments in a text-based simulation environment, we use the dataset from~\cite{huang2025limitlanguagemodelsplanning}.
Included are three simulated planning domains, BlocksWorld, Logistics, Barman from the International Planning Competition \cite{ipc}, with increasing action space and reported difficulty. We also consider Mystery BlocksWorld~\cite{valmeekam2023planbenchextensiblebenchmarkevaluating} where all keywords are perturbed to combat LLM memorization. Each instance comes with domain and problem descriptions and ground-truth PDDL domain and problem files that are used to validate a predicted plan. Each domain has 100 tasks of varying problem complexity and description naturalness. We use the heavily templated descriptions which are also the easiest due to the reported close-to-zero performance of LLMs with less than 100B parameters that we focus on. We crawl, process and use the Planning Wiki\footnote{\url{https://planning.wiki/guide/whatis/pddl}} as the source of documentation of the PDDL language.

\paragraph{Metrics}
We follow \citet{kagitha2025addressingchallengesplanninglanguage} and use syntactic and semantic accuracy to assess the DF and PF generated by an LLM. Syntactic accuracy is the percentage of problems where no syntax error are returned by the planning solver. Semantic accuracy is the percentage of problems where a plan is not only found but also correct. We use the dual-bfws-ffparser planner \citet{muise-icaps16demo-pd} to solve for the plan and the VAL4~\cite{1374201} to validate the plan against the gold DF and PF.

\paragraph{Model}
We conduct experiments on four open-source models, ranging from 8B to 32B parameters: Llama-4-Maverick-17B-128E-Instruct, Llama-4-Scout-17B-16E-Instruct\footnote{\url{https://github.com/meta-llama/llama-models/tree/main/models/llama4}}, QwQ-32B, Qwen3-8B\footnote{\url{https://github.com/QwenLM/Qwen3}}. We follow most cited previous works and only consider zero-shot prompting.





\begin{table*}[ht]
\renewcommand{\arraystretch}{0.8}
\small
\centering
\begin{tabular}{llccc}
\toprule
\textbf{Domain} & \textbf{Method} & \textbf{Metric} & \textbf{Qwen3-8b} & \textbf{Llama-4-Maverick} \\
\midrule
\multirow{4}{*}{Blocksworld} 
& \multirow{2}{*}{Feedback-Retrieved} 
& Syntax   & 41 / 42 ({\color{red}+1})  & 43 / 93 ({\color{red}+50}) \\
& & Semantic & 26 / 30 ({\color{red}+4})  & 39 / 85 ({\color{red}+46}) \\
& \multirow{2}{*}{Code-Retrieved} 
& Syntax   & 44 / 44 (0)  & 96 / 97 ({\color{red}+1}) \\
& & Semantic & 32 / 28 ({\color{blue}-4})  & 86 / 90 ({\color{red}+4}) \\
\midrule
\multirow{4}{*}{Mystery Blocksworld} 
& \multirow{2}{*}{Feedback-Retrieved} 
& Syntax   & 8 / 14 ({\color{red}+6})  & 35 / 67 ({\color{red}+32}) \\
& & Semantic & 0 / 0 (0) & 24 / 51 ({\color{red}+27}) \\
& \multirow{2}{*}{Code-Retrieved} 
& Syntax   & 22 / 7 ({\color{blue}-15})   & 56 / 60 ({\color{red}+4}) \\
& & Semantic & 0 / 0 (0)   & 49 / 47 ({\color{blue}-2}) \\
\midrule
\multirow{4}{*}{Logistic} 
& \multirow{2}{*}{Feedback-Retrieved} 
& Syntax   & 42 / 40 ({\color{blue}-2})  & 59 / 56 ({\color{blue}-3}) \\
& & Semantic & 10 / 12 ({\color{red}+2}) & 50 / 60 ({\color{red}+10}) \\
& \multirow{2}{*}{Code-Retrieved} 
& Syntax   & 43 / 34 ({\color{blue}-9})  & 63 / 33 ({\color{blue}-30}) \\
& & Semantic & 11 / 8 ({\color{blue}-3})  & 55 / 30 ({\color{blue}-25}) \\
\midrule
\multirow{4}{*}{Barman} 
& \multirow{2}{*}{Feedback-Retrieved} 
& Syntax   & 0 / 0 (0)  & 0 / 0 (0) \\
& & Semantic & 0 / 0 (0)  & 0 / 0 (0) \\
& \multirow{2}{*}{Code-Retrieved} 
& Syntax   & 1 / 0 ({\color{blue}-1})  & 1 / 2 ({\color{red}+1}) \\
& & Semantic & 0 / 0 (0)  & 0 / 0 (0) \\
\bottomrule
\end{tabular}
\caption{Comparison of BM25 vs Embedding-base retriever results across domains, methods, and models. Values are reported as \textbf{BM25 / Embedding ($\Delta$)}, where $\Delta = \text{Embedding} - \text{BM25}$.}
\label{tab:bm25_vs_emb}
\end{table*}

\section{Results}

We present the following key conclusions based on the results shown in the Figure~\ref{fig:Results}. 

\textbf{Documentation brings significant performance improvement.} 
On BlocksWorld, most LLMs under the Base setting perform close to zero accuracy, as observed in previous work. However, when equipped with appropriate documentation, they demonstrate a dramatic increase in their ability to generate valid PDDL. While the improvement depends on the LLM, Llama-4-Maverick sees a dramatic improvement of syntactic accuracy from 0\% to over 90\% and semantic accuracy of 0\% to over 80\% with the help of documentation but regardless of error refinement. Other originally zero-performing models such as Llama-4-Scout see an improvement of 50\% for syntactic and 30\% for semantic accuracy. On more challenging domains, absolute performance for all LLMs are thwarted, while documentation still greatly improves syntactic accuracy for many models. Overall, models that previously failed entirely begin to become functional as planning formalizers.

\textbf{Specific docs significantly reduces syntax errors.}  
Documentation proves effective in reducing syntax errors during both initial PDDL generation (\textit{Modular w/ Specific Doc}) and subsequent error-correction (\textit{Refinement w/ Code-Retrieved Doc}). This effect is especially evident in the case of Llama-4-Scout, which fails to generate any valid PDDL originally regardless of whether error correction is applied. Only when supported by relevant docs can it successfully generate valid PDDL, many of which leading to correct plans. Notably, using feedback to retrieve doc does not lead to consistent or significant performance gains, as the retrieved documents often fail to accurately correspond to the actual errors. This highlights that retrieval based on error codes is more effective in improving the accuracy of documentation retrieval.

\begin{figure*}[t!]
	\includegraphics[scale=0.53]{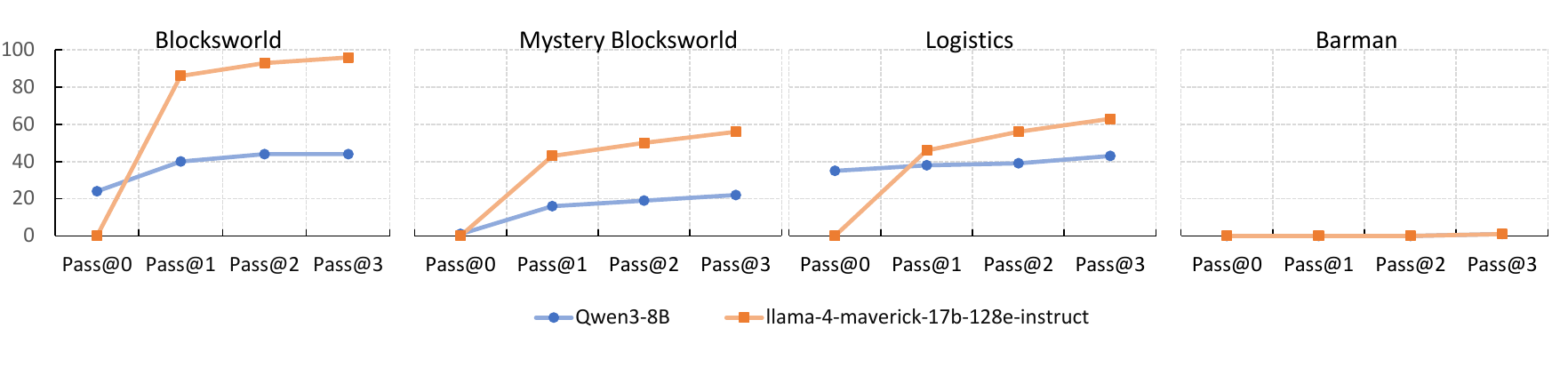}
	\caption{Syntactic accuracy on various rounds of Refinement w/ Code-Retrieved Doc.}
	\label{fig:pass}
\end{figure*}

\textbf{Docs cannot reliably reduce semantic errors.}  
During error correction, Llama-4-Maverick shows a 3\% improvement in syntax accuracy on the Logistic dataset under the \textit{Refinement w/ Code-Retrieved Doc} setting compared to the \textit{Refinement w/o Doc} setting. However, its semantic accuracy decreases by 1\%. This is because generating valid PDDL not only requires syntactic correctness but also an accurate representation of the environment. Otherwise, the resulting plan may fall into a loop, fail to reach the goal due to insufficient executable actions, or be unnecessarily complex. Achieving this depends heavily on the reasoning capabilities and world modeling abilities of the LLM, and simply providing documentation is not sufficient to enhance such reasoning.

\textbf{LLMs exhibit varying sensitivity to documentation across different phases of the code generation process.}
Our results reveal that documentation exerts a stronger influence during the initial code generation phase compared to the subsequent error refinement phase. Specifically, in the \textit{Formalize} phase—corresponding to the initial generation of PDDL—providing specific documentation significantly improves syntax accuracy, reaching up to 72\% for modular models with targeted documentation. In contrast, the benefits of documentation during the later \textit{Refinement} phase are substantially smaller. This suggests that models rely more on documentation cues when initially producing structured code, whereas later refinements depend more on internal representations and the code previously generated.
\begin{figure}[t!]
	\includegraphics[scale=0.25]{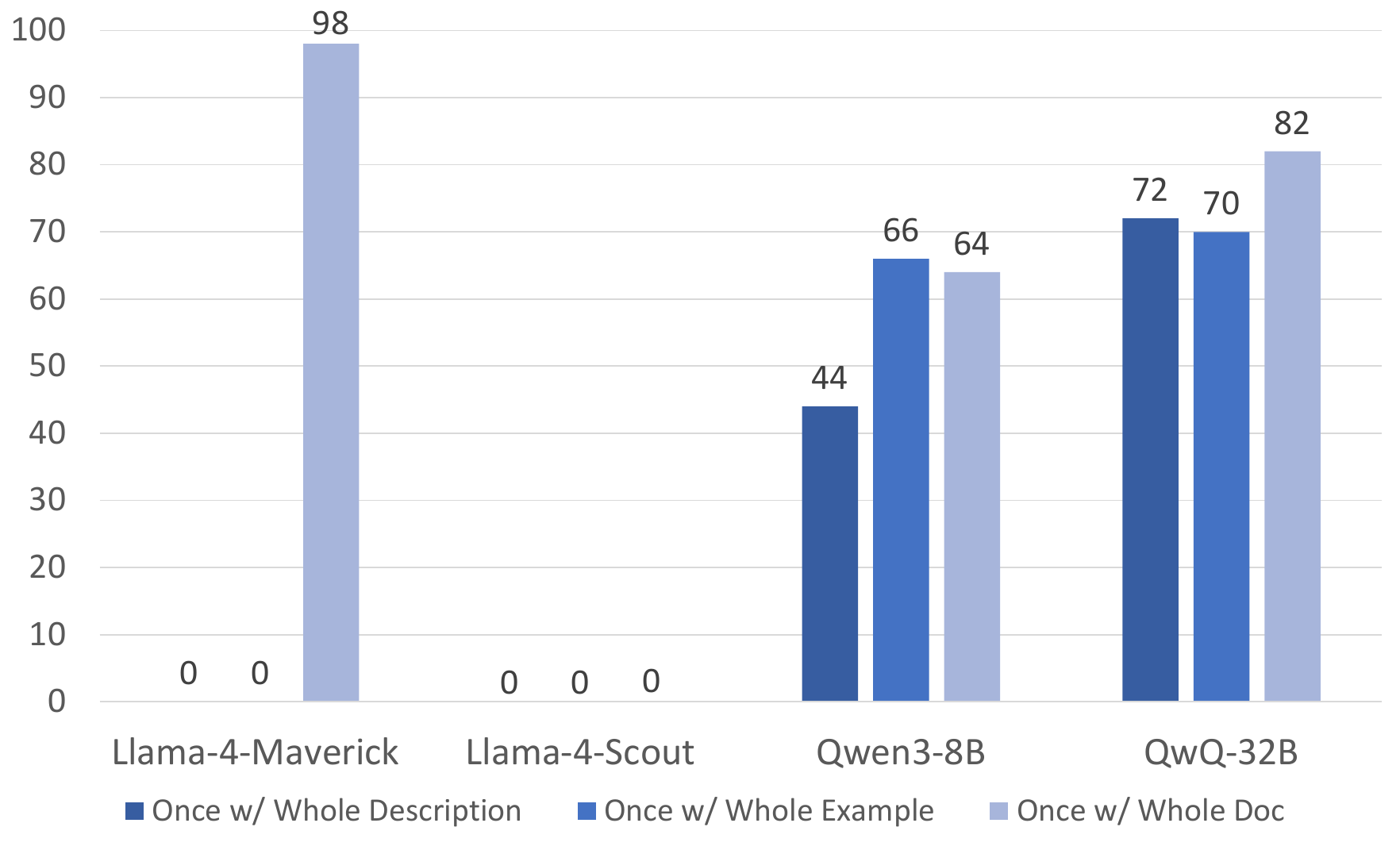}
	\caption{Syntactic accuracy of different models under various document conditions on BlocksWorld. Once w/ whole example refers to all the examples in the doc, and Once w/ whole description refers to all the textual descriptions in the doc.}
	\label{fig:Whole}
        \vspace{-0.5cm}
\end{figure}



\textbf{LLMs that are better at generating PDDL can make more effective use of documentation.}
Since QwQ-32B and Qwen3-8B outperform LLaMA-4 models in the \textit{Base} setting, we consider them more proficient at PDDL generation. Compared to the \textit{Base} and \textit{Modular w/ Specific Doc} settings, these PDDL-proficient models (QwQ-32B and Qwen3-8B) perform better under the \textit{Once w/ Whole Doc} setting. In contrast, the less proficient LLaMA-4 model does not outperform \textit{Modular w/ Specific Doc} under the same condition. This suggests that for models less capable of generating PDDL, modular generation is more effective, as they tend to become overwhelmed when processing large amounts of documentation.

\textbf{Using examples to convey knowledge is more effective than using descriptions.}  
Figure~\ref{fig:Whole} presents the performance of different types of documentation in the LLM-as-Formalizer setting. Among all types, \textit{Once w/ whole doc} yields the best results. Notably, for Llama-4-Maverick, performance is 0\% when provided with only examples or only descriptions, but nearly 100\% when given the entire documentation. Comparing \textit{Once w/ whole example} and \textit{Once w/ whole description}, we observe that examples consistently outperform descriptions. This suggests that examples are easier for LLMs to comprehend and are more useful for correcting syntax errors. Furthermore, even for models with inherently strong PDDL generation capabilities, such as QwQ-32B, the use of documentation still leads to a noticeable improvement in performance.

\textbf{The embedding-based retriever exhibits divergent effects across refinement settings.}
Table~\ref{tab:bm25_vs_emb} showing that in Refinement w/ Feedback-Retrieved Doc, replacing BM25 with text-embedding-3-small leads to substantial performance gains. For instance, llama-4-maverick-17b achieves 93\% syntax and 85\% semantic accuracy in Blocksworld, indicating that embeddings provide more precise retrieval guidance than BM25 in this context. Conversely, in Refinement w/ Code-Retrieved Doc, embeddings negatively impact performance. In the Logistic domain, llama-4-maverick-17b-128e-instruct drops to 33\% syntax accuracy compared to 63\% with BM25, while Qwen3-8b falls to 34\% compared to 43\% with BM25. In other domains, results remain roughly comparable to BM25. Overall, BM25 continues to achieve the strongest results for code-retrieved refinement, highlighting its robustness in this setting.

\textbf{Refinement yields substantial but diminishing improvements, with most gains concentrated in the first iteration.}As shown in Figure~\ref{fig:pass}. We evaluate refinement across 0--3 iterations on four benchmarks (Blocksworld, Mystery\_Blocksworld, Logistic, Barman). Results show that, starting from the 0-round baseline, refinement consistently improves performance, with the largest gains observed between 0~$\rightarrow$~1 round. For example, in Blocksworld, Qwen-8B improves from 24~$\rightarrow$~44 and llama-4-maverick-17b from 0~$\rightarrow$~96 in syntax accuracy after the first round. Beyond two rounds, the marginal improvements diminish, suggesting that a small number of refinement iterations is sufficient.

\section{Conclusion}
Our experiments clearly demonstrate that incorporating documentation to the process greatly improves generation of low-resource formal languages like PDDL. We show that for models less skilled at generating PDDL, documentation is only useful when paired with techniques like modular generation or error refinement. For more capable models, documentation accuracy matters more. Despite the clear gain, models still struggle when their size is small and when the domain is complex, which future work should strive to address. 


\section{Limitations}
While our proposed pipelines significantly improve the syntactic and, to a lesser extent, semantic accuracy of PDDL generation in low-resource settings, several limitations remain. First, our methods rely on well-structured documentation and domain descriptions; performance may degrade in noisy or under-specified environments. Moreover, documentation itself may contain outdated, incomplete, or inaccurate information, which can mislead the model during generation. Second, although documentation helps reduce syntax errors, semantic correctness still heavily depends on the model's internal reasoning capabilities, which are limited for smaller LLMs. Lastly, our evaluation is confined to a few benchmark domains; generalization to more diverse or real-world planning scenarios remains to be verified.

The datasets we use are all under the MIT License.

\bibliography{anthology,custom}

\appendix

\section{Data and PDDL Examples}
\label{sec:example}
\begin{figure*}[htb]
	\includegraphics[scale=0.85]{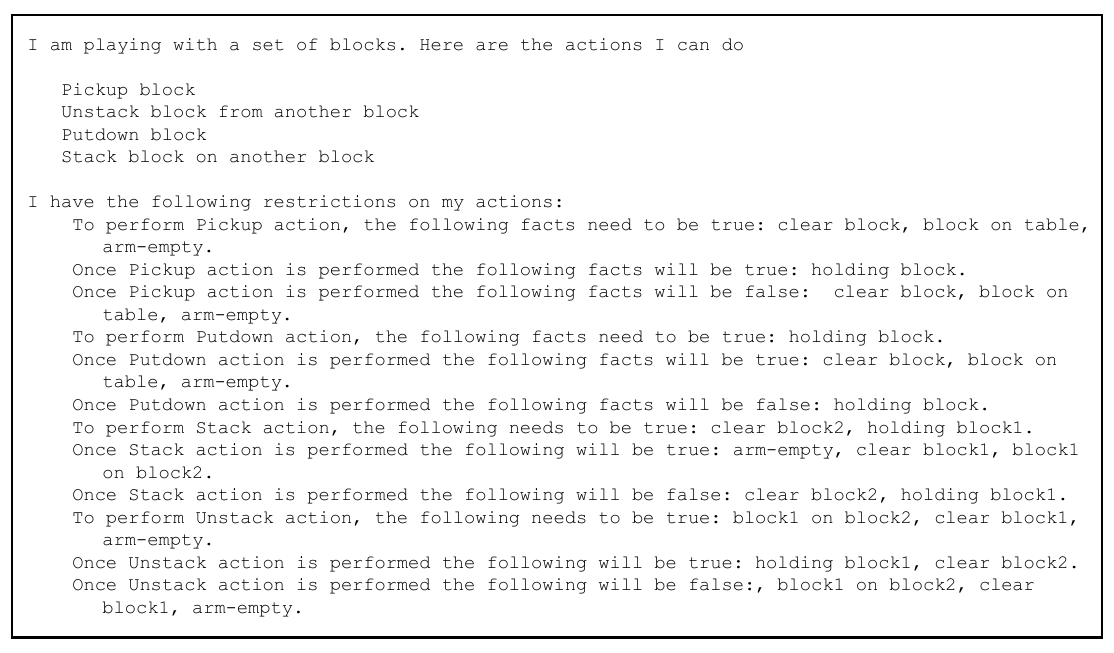}
	\caption{DD for the BlocksWorld domain}
	\label{fig:DD}
\end{figure*}

\begin{figure*}[htb]
	\includegraphics[scale=0.85]{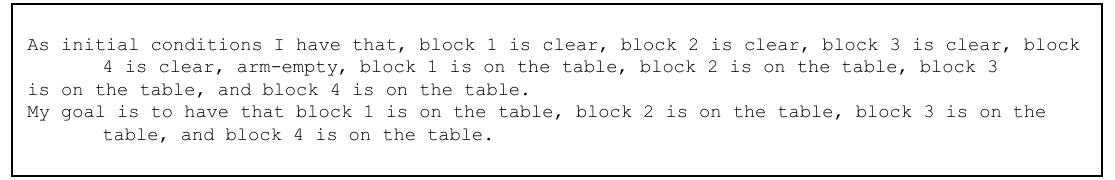}
	\caption{PD for the BlocksWorld domain}
	\label{fig:PD}
\end{figure*}

Figure~\ref{fig:DD} and~\ref{fig:PD} is an example of the dataset Heavily\_Templated\_BlocksWorld-100 from~\cite{huang2025limitlanguagemodelsplanning}.

\begin{figure*}[htb]
	\includegraphics[scale=0.85]{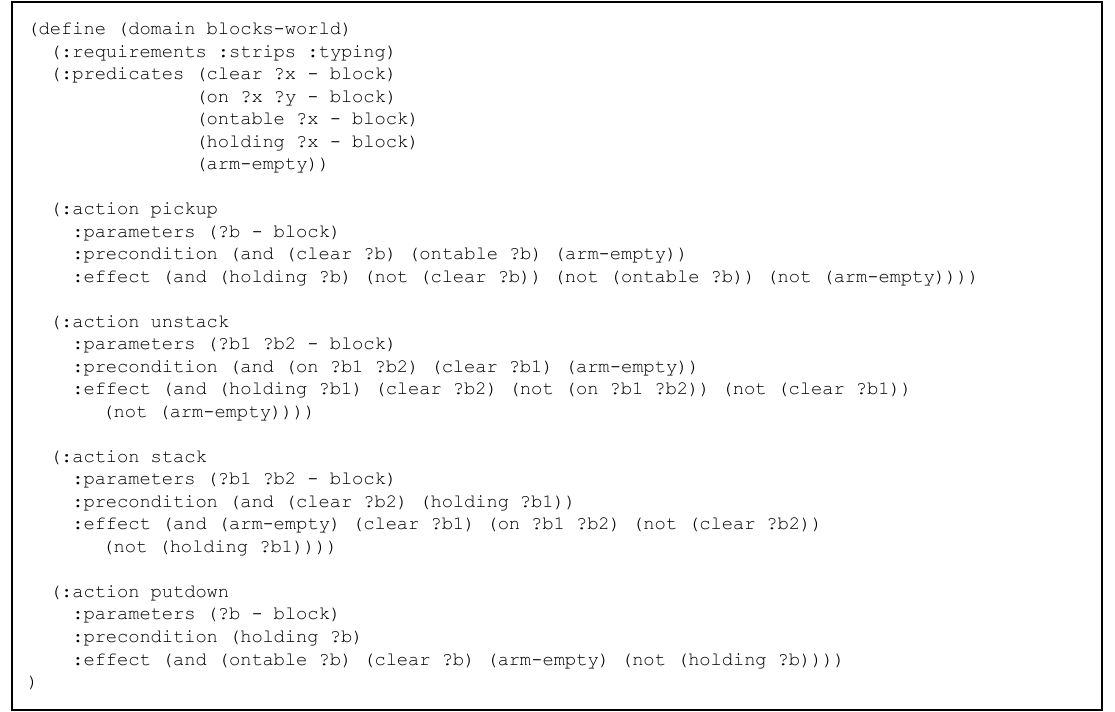}
	\caption{DF for the BlocksWorld domain}
	\label{fig:DF}
\end{figure*}

\begin{figure*}[htb]
	\includegraphics[scale=0.85]{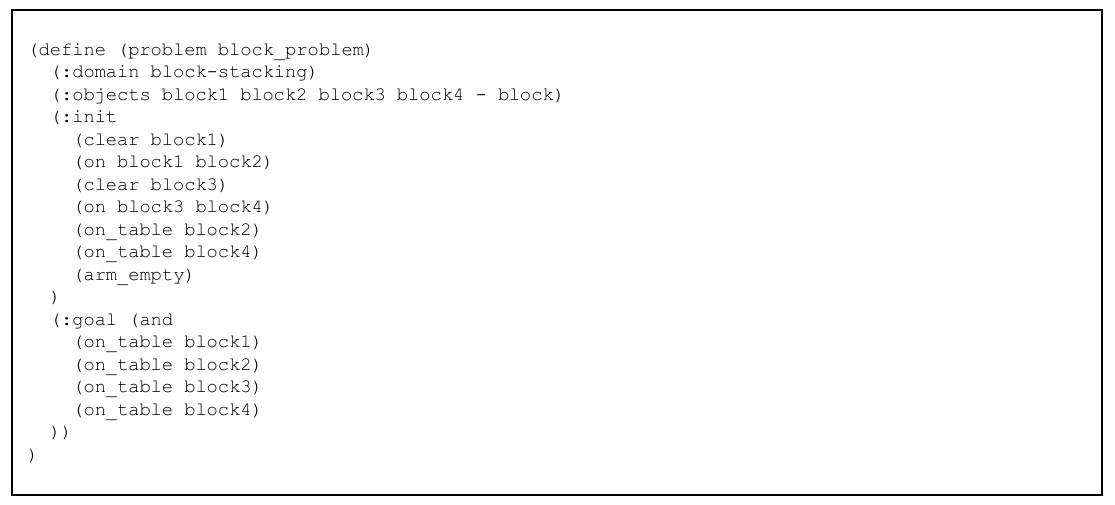}
	\caption{PF for the BlocksWorld domain}
	\label{fig:PF}
\end{figure*}

\section{Z3 Result}
\label{sec:Z3}
We followed the~\cite{hao2025planningrigorgeneralpurposezeroshot} by using Formulator to define all possible variables in the environment and generate their instantiation information before producing the Z3 code. However, we did not adopt their iterative error correction method. In their experiments, Formulator improved the results on the BlocksWorld domain from 0.2 to 96.2.  

We conducted experiments on our dataset using GPT-4o as the LLM, but the results were 0. The distribution of error causes is shown in the Table~\ref{tab:Z3}. Goal unsatisfied means that the final output plan cannot solve the problem correctly. We analyzed the cause of this error. We printed the state of each time slice and found that as long as any condition in the goal state is met, the planning will stop. When we tried to let LLM correct this error, it only caused more syntax errors, and never corrected the error. This is likely because our dataset is more complex—theirs only involved 4 blocks, whereas ours often includes more than 10 blocks.  

Since even the simplest BlocksWorld dataset yielded a score of 0 after following the~\cite{hao2025planningrigorgeneralpurposezeroshot} approach, we did not apply our pipeline to Z3 and instead reported the findings in the appendix.
\begin{table}[htb]
    \centering
    \begin{tabular}{lcc}
    \toprule
        & \multicolumn{2}{c}{Heavily BlocksWorld}\\
        Model & syntax error &  goal unsatisfied\\
         \toprule[0.3mm]
        gpt-4o & 16/100 & 84/100\\
        
        \toprule[0.3mm]

    \end{tabular}
    \caption{Z3 Result}
    \label{tab:Z3}
    \vspace{-0.5cm}
\end{table}

\section{Pseudocode of Refinement w/ Code-Retrieved Doc}
\label{sec:pseudocode}
Algorithm~\ref{alg:rag_pddl} shows the Pseudocode of Refinement w/ Code-Retrieved Doc.
\begin{algorithm}[htb]
\caption{Retrieval-Augmented PDDL Generation with Iterative Correction}
\label{alg:rag_pddl}
\begin{algorithmic}[1]
\REQUIRE Domain Description (DD), Problem Description (PD)
\ENSURE Valid Domain File (DF) and Problem File (PF)

\STATE $\langle DF, PF \rangle \leftarrow \text{LLM}(DD, PD)$

\WHILE{true}
    \STATE $feedback \leftarrow \text{Solver}(DF, PF)$
    \IF{feedback indicates success}
        \RETURN $\langle DF, PF \rangle$
    \ENDIF

    \STATE $e\_type \leftarrow \text{Parse\_Error\_Type}(feedback)$

    \IF{$e\_type$ == \texttt{syntax\_error} \AND $feedback.file$ == \texttt{DF}}
        \STATE $e\_code \leftarrow \text{LLM}(feedback)$
        \STATE $doc \leftarrow \text{Retrieve}(e\_code)$
        \STATE $\langle DF, PF \rangle \leftarrow \text{LLM}(DF, PF, e\_code, feedback, doc)$

    \ELSIF{$e\_type$ == \texttt{syntax\_error} \AND $feedback.file$ == \texttt{PF}}
        \STATE $\langle DF, PF \rangle \leftarrow \text{LLM}(DF, PF, feedback)$

    \ELSIF{$e\_type$ == \texttt{semantic\_error}}
        \STATE $\langle DF, PF \rangle \leftarrow \text{LLM}(DF, PF, feedback)$

    \ELSE
        \STATE \textbf{raise} \texttt{UnknownErrorType}
    \ENDIF
\ENDWHILE
\end{algorithmic}
\end{algorithm}

\section{PDDL Error Cases and Corrections}
\label{sec:Error case}
\subsection{Example 1: Action Definition Error}

We use \texttt{@@@ ... @@@} to clearly mark errors. In the original definition, the action was generated as:

\begin{lstlisting}[language=Lisp]
(:action pickup 
 :parameters (?b)

 @@@:preconditions@@@ (and (clear ?b) 
                           (on-table ?b) 
                           (arm-empty))

 :effects (and (holding ?b) 
               (not (clear ?b)) 
               (not (on-table ?b)) 
               (not (arm-empty))))
\end{lstlisting}

In PDDL, \texttt{:precondition} must be strictly singular. Therefore, the solver returns the error message:  
\texttt{domain: syntax error in line 12, ':PRECONDITIONS': domain definition expected}.

Based on this error, BM25 retrieved the following documentation:

\begin{quote}
\textbf{type\_name:} Actions  

\textbf{documentation:} An action defines a transformation in the state of the world. It is broken down into three sections:  

1. \texttt{:parameters} — entities involved in the action.  
2. \texttt{:precondition} — conditions required for applicability.  
3. A choice between \texttt{:effect} and \texttt{:expansion} (most domains use \texttt{:effect}).  

\textbf{Example:}  
\begin{lstlisting}[language=Lisp]
(:action BUILD-WALL
 :parameters (?s - site ?b - bricks)
 :precondition (and (on-site ?b ?s) 
                    (foundations-set ?s) 
                    (not (walls-built ?s)) 
                    (not (material-used ?b)))
 :effect (and (walls-built ?s) 
              (material-used ?b)))
\end{lstlisting}
\end{quote}

The corrected PDDL definition is:

\begin{lstlisting}[language=Lisp]
(:action pickup
 :parameters (?b)
 :precondition (and (clear ?b) 
                    (on-table ?b) 
                    (arm-empty))
 :effects (and (holding ?b) 
               (not (clear ?b)) 
               (not (on-table ?b)) 
               (not (arm-empty))))
\end{lstlisting}

\subsection{Example 2: Predicate Definition Error}

In the original definition, the predicates were generated as:

\begin{lstlisting}[language=Lisp]
(:predicates
 (on-table ?obj - container)
 (hand-empty ?hand - hand)
 (holding ?hand - hand ?container - container)
 (dispenses ?dispenser - dispenser ?ingredient - ingredient)
 (empty ?container - container)
 (clean ?container - container)
 (used-with ?container - container 
            ?item - @@@(ingredient cocktail)@@@))
\end{lstlisting}

In PDDL, each parameter can only be assigned a single type. Therefore, the solver returns the error message:  
\texttt{domain: syntax error in line 14, '(': domain definition expected}.

BM25 retrieved the following documentation:

\begin{quote}
\textbf{type\_name:} Predicates  

\textbf{documentation:} Predicates represent the state of the system in PDDL and can be either true or false at any given moment. They usually apply to specific types of objects and can take one or more arguments.  

\textbf{Example:}  
\begin{lstlisting}[language=Lisp]
(:predicates
 (walls-built ?s - site)
 (windows-fitted ?s - site)
 (foundations-set ?s - site)
 (cables-installed ?s - site)
 (site-built ?s - site)
 (on-site ?m - material ?s - site)
 (material-used ?m - material))
\end{lstlisting}
\end{quote}

The corrected PDDL definition is:

\begin{lstlisting}[language=Lisp]
(:predicates
 (on-table ?obj)
 (hand-empty ?hand)
 (holding ?hand ?container)
 (dispenses ?dispenser ?ingredient)
 (empty ?container)
 (clean ?container)
 (used-with ?container ?item))
\end{lstlisting}

\section{Prompt}
\label{sec:prompt}

\begin{figure*}[htb]
	\includegraphics[scale=0.60]{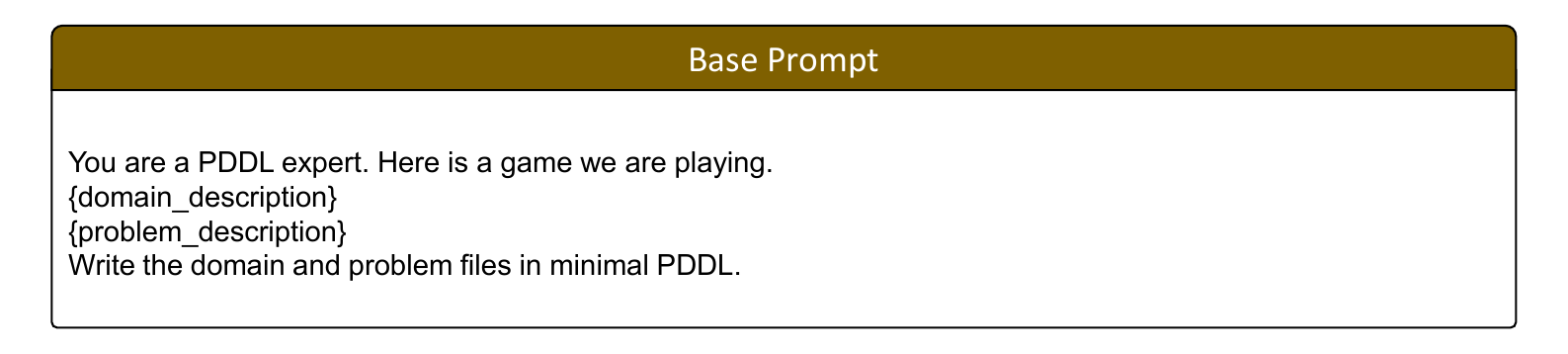}
	\caption{Base Prompt}
	\label{fig:base}
\end{figure*}

\begin{figure*}[htb]
	\includegraphics[scale=0.60]{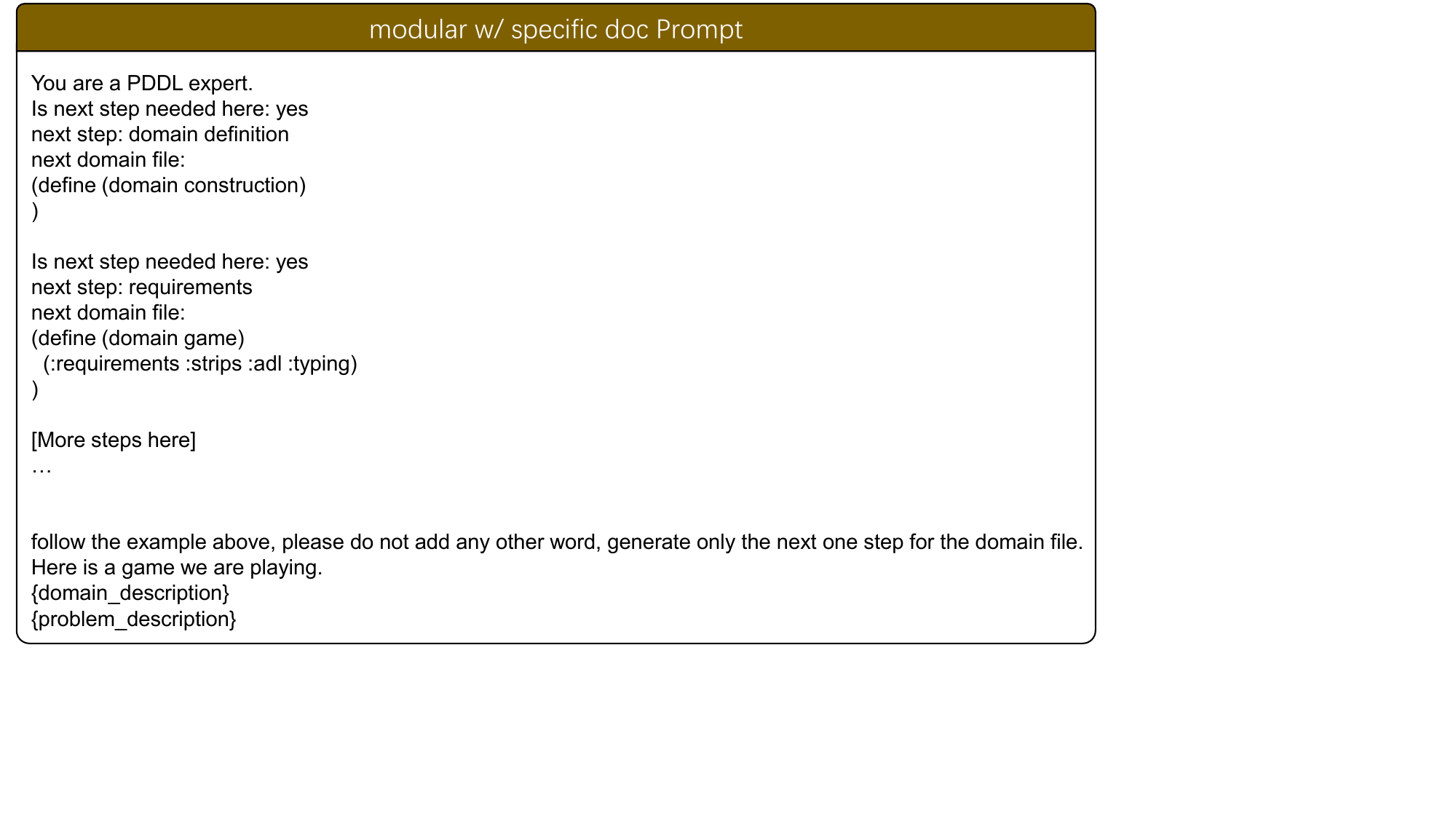}
	\caption{modular w/ specific doc Prompt}
	\label{fig:modular}
\end{figure*}

\begin{figure*}[htb]
	\includegraphics[scale=0.60]{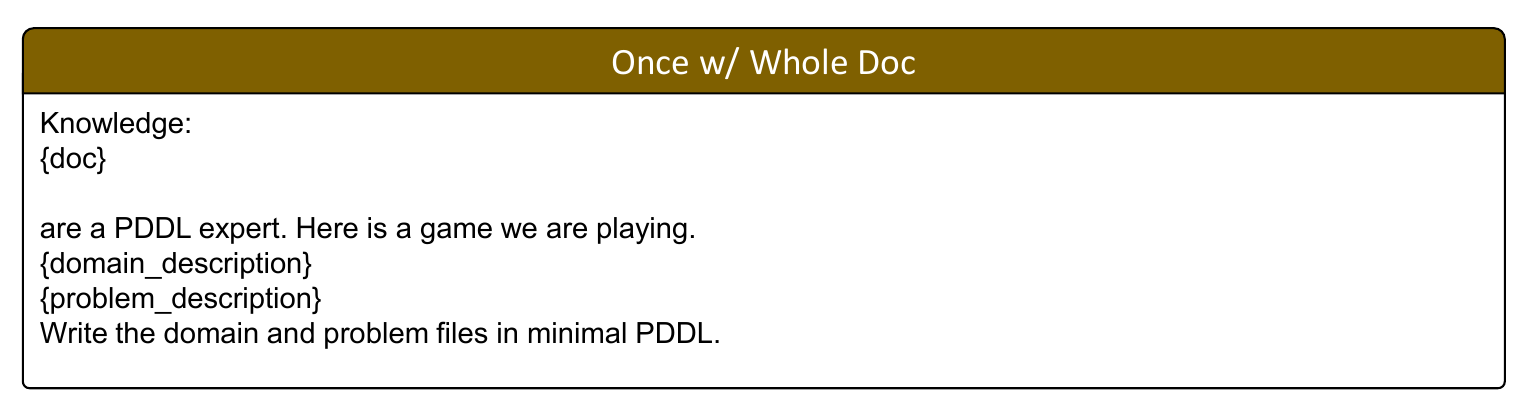}
	\caption{Once w/ Whole Doc Prompt}
	\label{fig:whole prompt}
\end{figure*}

\begin{figure*}[htb]
	\includegraphics[scale=0.60]{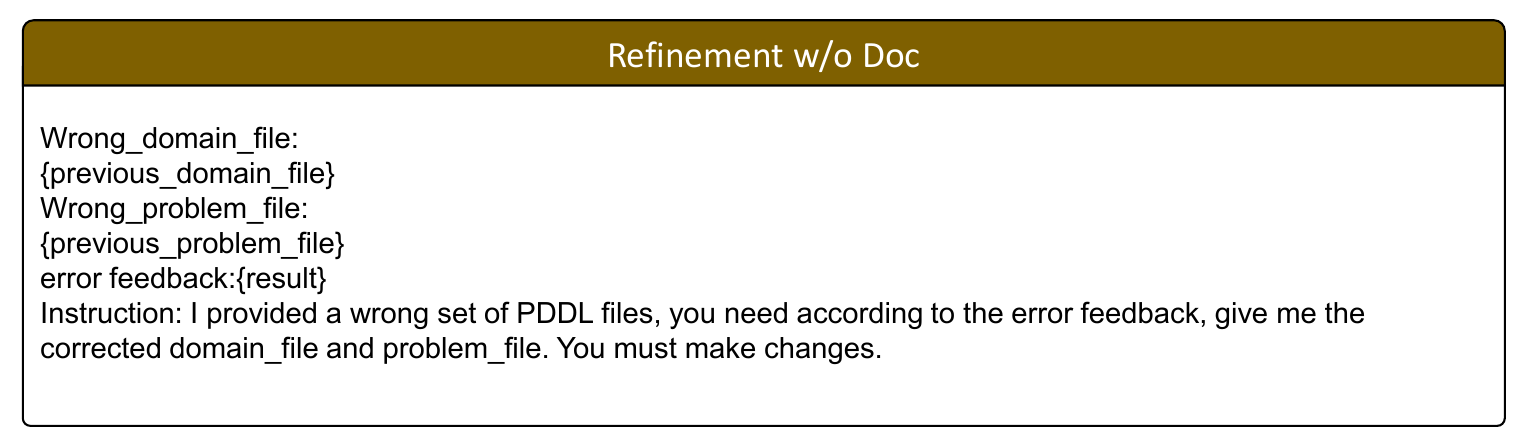}
	\caption{Refinement w/o Doc Prompt}
	\label{fig:w/ error}
\end{figure*}

\begin{figure*}[htb]
	\includegraphics[scale=0.60]{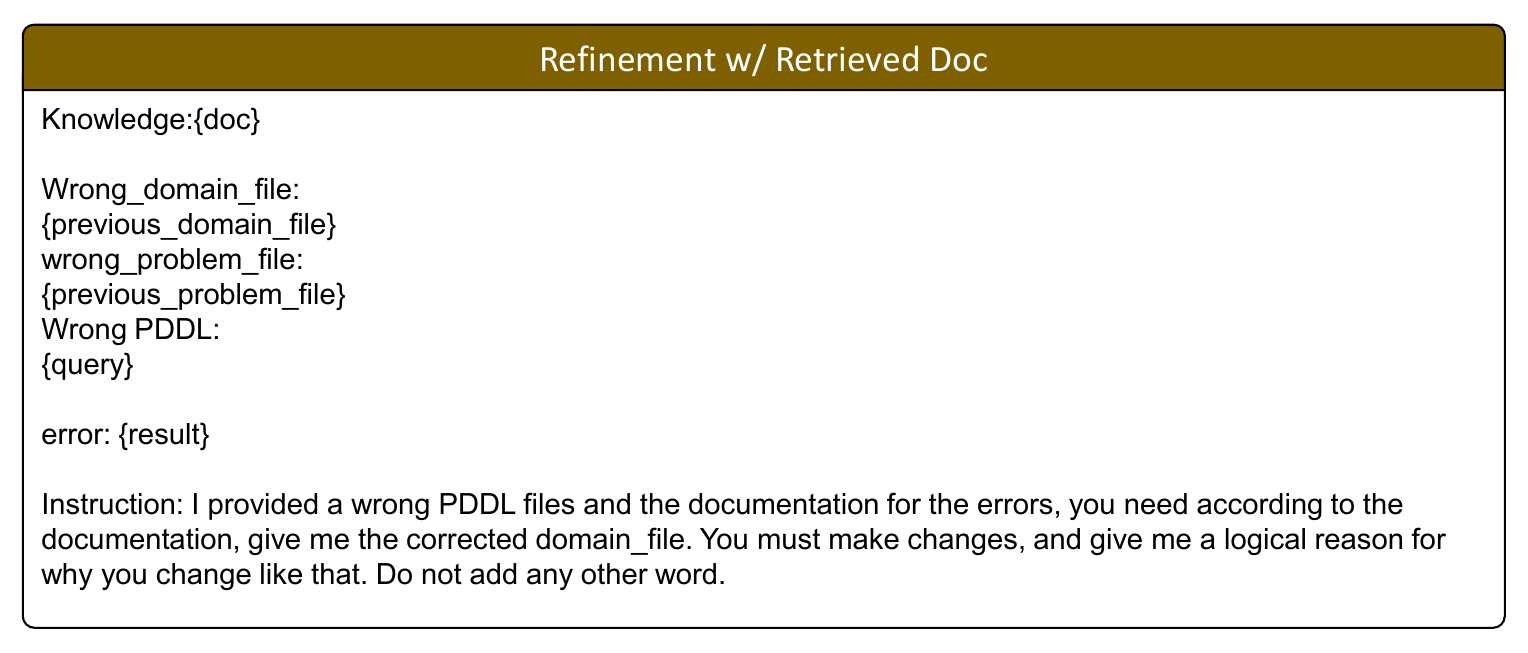}
	\caption{Refinement w/ Retrieved Doc Prompt}
	\label{fig:w/ retrieved}
\end{figure*}

Figure~\ref{fig:base}~\ref{fig:modular}~\ref{fig:whole prompt}~\ref{fig:w/ error} and~\ref{fig:w/ retrieved} is the Prompt of all our methods. Refinement w/ Feedback-Retrieved Doc and Refinement w/ Code-Retrieved Doc use the same prompt but different retrieved docs.

\end{document}